\documentclass[12pt]{article}

\newcommand{\text}{\rm}

\begin{document}

\title{ \textbf{On the $SL(2,R)$ symmetry in
Yang-Mills Theories in the Landau, Curci-Ferrari and Maximal Abelian Gauge } 
 }
\author{D. Dudal\thanks{%
Research Assistant of The Fund For Scientific Research-Flanders, Belgium.},\
\ H. Verschelde \thanks{%
David.Dudal@rug.ac.be, henri.verschelde@rug.ac.be} \\
{\small {\textit{Ghent University }}}\\
{\small {\textit{Department of Mathematical Physics and Astronomy,
Krijgslaan 281-S9, }}}\\
{\small {\textit{B-9000 GENT, BELGIUM}}} \and V.E.R. Lemes, M.S. Sarandy,
and S.P. Sorella\thanks{%
vitor@dft.if.uerj.br, sarandy@dft.if.uerj.br, sorella@uerj.br} \\
{\small {\textit{UERJ - Universidade do Estado do Rio de Janeiro,}}} \\
{\small {\textit{\ Rua S\~{a}o Francisco Xavier 524, 20550-013 Maracan\~{a}, 
}}} {\small {\textit{Rio de Janeiro, Brazil.}}} \and M. Picariello\thanks{%
Marco.Picariello@mi.infn.it }, \\
{\small {\textit{Universit\'{a} degli Studi di Milano, via Celoria 16,
I-20133, Milano, Italy }}}\\
{\small {\textit{and INFN\ Milano, Italy}}}}
\maketitle

\begin{abstract}
 The existence of a $SL(2,R)$ symmetry is
discussed in $SU(N)$ Yang-Mills in the maximal Abelian Gauge. This symmetry,
also present in the Landau and Curci-Ferrari gauge, ensures the absence of
tachyons in the maximal Abelian gauge. In all these gauges, \emph{$SL(2,R)$ }%
turns out to be dynamically broken by ghost condensates. 
\end{abstract}

\vfill\newpage\ \makeatother

\renewcommand{\theequation}{\thesection.\arabic{equation}}

\section{Introduction}

It is widely believed that the dual superconductivity mechanism \cite
{scon,th} can be at the origin of color confinement. The key ingredients of
this mechanism are the Abelian dominance and the monopoles condensation.
According to the dual superconductivity picture, the low energy behavior of $%
QCD$ should be described by an effective Abelian theory in the presence of
monopoles. The condensation of the monopoles gives rise to the formation of
Abrikosov-Nielsen-Olesen flux tubes which confine all chromoelectric
charges. This mechanism has received many confirmations from lattice
simulations in Abelian gauges, which are very useful in order to
characterize the effective relevant degrees of freedom at low energies.

Among the Abelian gauges, the so called maximal Abelian gauge (MAG) plays an
important role. This gauge, introduced in \cite{th,ks}, has given evidences
for the Abelian dominance and for the monopoles condensation, while
providing a renormalizable gauge in the continuum. Here, the Abelian degrees
of freedom are identified with the components of the gauge field belonging
to the Cartan subgroup of the gauge group $SU(N)$. The other components
correspond to the $\left( N^{2}-N\right) $ off-diagonal generators of $SU(N)$
and, being no longer protected by gauge invariance, are expected to acquire
a mass, thus decoupling at low energies. The understanding of the mechanism
for the dynamical mass generation of the off-diagonal components is
fundamental for the Abelian dominance.

A feature to be underlined is that the MAG is a nonlinear gauge. As a
consequence, a quartic self-interaction term in the Faddeev-Popov ghosts is
necessarily required for renormalizability \cite{mp,fz}. Furthermore, as
discussed in \cite{ms,k} and later on in \cite{sp}, the four ghost
interaction gives rise to an effective potential whose vacuum configuration
favors the formation of off-diagonal ghost condensates $\left\langle c%
\overline{c}\right\rangle $, $\left\langle cc\right\rangle $ and $%
\left\langle \overline{c}\overline{c}\right\rangle \footnote{%
The notation for the ghost condensates $\left\langle c\overline{c}%
\right\rangle $, $\left\langle cc\right\rangle $ and $\left\langle \overline{%
c}\overline{c}\right\rangle $ stands for $\left\langle f^{iab}c^{a}\overline{%
c}^{b}\right\rangle $, $\left\langle f^{iab}c^{a}c^{b}\right\rangle $ and $%
\left\langle f^{iab}\overline{c}^{a}\overline{c}^{b}\right\rangle $, where $%
f^{iab}$ are the structure constants of the gauge group. The index $i$ runs
over the Cartan generators, while the indices $a,b$ correspond to the
off-diagonal generators.}$.\ However, these ghost condensates were proven 
\cite{dd} to originate an unwanted effective tachyon mass for the
off-diagonal gluons, due to the presence in the MAG of an off-diagonal
interaction term of the type $AAc\overline{c}$.

Meanwhile, the ghost condensation has been observed in others gauges, namely
in the Curci-Ferrari gauge \cite{sp2} and in the Landau gauge \cite{sp3}. We
remark that in these gauges the ghost condensates do not give rise to any
mass term for the gauge fields. The existence of these condensates turns out
to be related to the dynamical breaking of a $SL(2,R)$ symmetry which is
known to be present in both Curci-Ferrari and Landau gauge since long time 
\cite{no,ni,dj,oji}. It is worth noticing that in the Curci-Ferrari gauge a
BRST invariant mass term $\left( \frac{1}{2}\mathcal{A}^{2}-\xi c\overline{c}%
\right) \equiv \left( \frac{1}{2}\mathcal{A}^{A}\mathcal{A}^{A}-\xi c^{A}%
\overline{c}^{A}\right) $, with $A=1,...,N^{2}-1\;$and $\xi $ being the
gauge parameter, can be introduced without spoiling the renormalizability of
the model \cite{cf,ds}.

Recent investigations \cite{ope} have suggested that the existence of a
nonvanishing condensate $\left( \frac{1}{2}\left\langle \mathcal{A}%
^{2}\right\rangle -\xi \left\langle c\overline{c}\right\rangle \right) $
could yield a BRST invariant dynamical mass for both gluons and ghosts. We
observe that in the limit $\xi \rightarrow 0$, this condensate reduces to
the pure gauge condensate $\frac{1}{2}\left\langle \mathcal{A}%
^{2}\right\rangle $ whose existence is well established in the Landau gauge 
\cite{h1,pb}, providing indeed an effective mass for the gluons.

The aim of this paper is to show that the $SL(2,R)$ symmetry is also present
in the MAG for $SU(N)$ Yang-Mills, with any value of $N$. This result
generalizes that of \cite{ms,k,sawa}, where the $SL(2,R)$ symmetry has been
established only for a partially gauge-fixed version of the action. In
particular, the requirement of an exact $SL(2,R)$ invariance for the
complete quantized action, including the diagonal part of the gauge fixing,
will have very welcome consequences. In fact, as we shall see, this
requirement introduces new interaction terms in the action, which precisely
cancel the term $AAc\overline{c}$ responsible for the generation of the
tachyon mass. In other words, no tachyons are present if the $SL(2,R)$
symmetry is required from the beginning as an exact invariance of the fully
gauge-fixed action.

This observation allows us to make an interesting {\textit{trait d' union }}
between the Landau gauge, the Curci-Ferrari gauge and the MAG, providing a
more consistent and general understanding of the ghost condensation and of
the mechanism for the dynamical generation of the effective gluon masses.
The whole framework can be summarized as follows. The ghost condensates
signal the dynamical breaking of the $SL(2,R)\;$symmetry, present in all
these gauges. Also, the condensed vacuum has the interesting property of
leaving unbroken the Cartan subgroup of the gauge group. As a consequence of
the ghost condensation, the off-diagonal ghost propagators get deeply
modified in the infrared region \cite{ms,k,sp,sp3}. This feature might be
relevant for the analysis of the infrared behavior of the gluon propagator,
through the ghost-gluon mixed Schwinger-Dyson equations. Moreover, the ghost
condensates contribute to the dimension four condensate $\left\langle \frac{%
\alpha }{\pi }F^{2}\right\rangle $ through the trace anomaly.

On the other hand, the dynamical mass generation for the gluons is expected
to be related to the BRST invariant condensate $\left( \frac{1}{2}%
\left\langle \mathcal{A}^{2}\right\rangle -\xi \left\langle c\overline{c}%
\right\rangle \right) $. It is remarkable that this condensate can be
defined in a BRST invariant way also in the MAG \cite{ope,dd}, where it can
give masses for all off-diagonal fields, thus playing a pivotal role for the
Abelian dominance.

The paper is organized as follows. Sect.2 is devoted to the analysis of the $%
SL(2,R)$ symmetry in the Landau, Curci-Ferrari and MAG gauge. In Sect.3 we
prove that the requirement of the $SL(2,R)$ symmetry for the MAG in $SU(N)$
Yang-Mills yields a renormalizable theory. In Sect.4 we discuss the issue of
the ghost condensation and of the absence of tachyons. In Sect.5 we present
the conclusions.

\section{Yang-Mills theories and the $SL(2,R)$ symmetry}

Let $\mathcal{A}_{\mu }$ be the Lie algebra valued connection for the gauge
group $SU(N),$ whose generators $T^{A}\,,\;\left[ T^{A},T^{B}\right]
=f^{ABC}T^{C}\,$, are chosen to be antihermitean and to obey the
orthonormality condition $\mathrm{Tr}\left( T^{A}T^{B}\right) =\delta ^{AB}$%
, with $A,B,C=1,..,\left( N^{2}-1\right) $. The covariant derivative is
given by 
\begin{equation}
\mathcal{D}_{\mu }^{AB}\equiv \partial _{\mu }\delta ^{AB}-gf^{ABC}\mathcal{A%
}_{\mu }^{C}\;.  \label{dudal0}
\end{equation}
Let $s$ and $\overline{s}$ be the nilpotent BRST and anti-BRST
transformations respectively, acting on the fields as 
\begin{eqnarray}
s\mathcal{A}_{\mu }^{A} &=&-\mathcal{D}_{\mu }^{AB}c^{B}  \nonumber \\
sc^{A} &=&\frac{g}{2}f^{ABC}c^{B}c^{C}  \nonumber \\
s\overline{c}^{A} &=&b^{A}  \nonumber \\
sb^{A} &=&0\;,  \label{s}
\end{eqnarray}
\begin{eqnarray}
\overline{s}\mathcal{A}_{\mu }^{A} &=&-\mathcal{D}_{\mu }^{AB}\overline{c}%
^{B}  \nonumber \\
\overline{s}c^{A} &=&-b^{A}+gf^{ABC}c^{B}\overline{c}^{C}  \nonumber \\
\overline{s}\;\overline{c}^{A} &=&\frac{g}{2}f^{ABC}\overline{c}^{B}%
\overline{c}^{C}  \nonumber \\
\overline{s}b^{A} &=&-gf^{ABC}b^{B}\overline{c}^{C}\;.  \label{as}
\end{eqnarray}
Here $c^{A}$ and $\overline{c}^{A}$ generally denote the Faddeev-Popov
ghosts and anti-ghosts, while $b^{A}$ denote the Lagrange multipliers.

Furthermore, we define the operators $\delta $ and $\overline{\delta }$ by 
\begin{eqnarray}
\delta \overline{c}^{A} &=&c^{A}  \nonumber \\
\delta b^{A} &=&\frac{g}{2}f^{ABC}c^{B}c^{C}  \nonumber \\
\delta \mathcal{A}_{\mu }^{A} &=&\delta c^{A}=0\;,  \label{df}
\end{eqnarray}
\begin{eqnarray}
\overline{\delta }c^{A} &=&\overline{c}^{A}  \nonumber \\
\overline{\delta }b^{A} &=&\frac{g}{2}f^{ABC}\overline{c}^{B}\overline{c}^{C}
\nonumber \\
\overline{\delta }\mathcal{A}_{\mu }^{A} &=&\overline{\delta }\overline{c}%
^{A}=0\;.  \label{adf}
\end{eqnarray}
Together with the Faddeev-Popov ghost number operator $\delta _{FP}$, $%
\delta $ and $\overline{\delta }$ generate a $SL(2,R)$ algebra. This algebra
is a subalgebra of the algebra generated by $\delta _{FP}$, $\delta $, $%
\overline{\delta }$ and the BRST and anti-BRST operators $s$ and $\overline{s%
}$. The algebra 
\begin{eqnarray}
s^{2} &=&0\,,\,\,\,\,\,\,\,\,\,\,\,\overline{s}^{2}=0\;,  \nonumber
\label{ojima} \\
\,\{s,\overline{s}\} &=&0\,,\,\,\,\,\,\,\,\,\,\,[\delta ,\overline{\delta }%
]=\delta _{\mathrm{FP}}\,,  \nonumber \\
\lbrack \delta ,\delta _{\mathrm{FP}}] &=&-2\delta ,\,\,\,[\overline{\delta }%
,\delta _{\mathrm{FP}}]=2\overline{\delta }\;,  \nonumber \\
\lbrack s,\delta _{\mathrm{FP}}] &=&-s\,,\,\,\,\,\,\,[\overline{s},\delta _{%
\mathrm{FP}}]=\overline{s}\,,  \nonumber \\
\lbrack s,\delta ] &=&0\,\,,\,\,\,\,\,\,\,\,[\overline{s},\overline{\delta }%
]=0\,,  \nonumber \\
\lbrack s,\overline{\delta }] &=&-\overline{s}\,\,,\,\,\,\,\,[\overline{s}%
,\delta ]=-s\,,  \label{comm}
\end{eqnarray}
is known as the Nakanishi-Ojima (NO) algebra \cite{no}.

\subsection{Landau Gauge}

In the Landau gauge, we have 
\begin{equation}
S=S_{YM}+S_{GF+FP}=-\frac{1}{4}\int d^{4}x\mathcal{F}_{\mu \nu }^{A}\mathcal{%
F}^{A\mu \nu }+s\int d^{4}x\overline{c}^{A}\partial _{\mu }\mathcal{A}^{A\mu
}\;.  \label{dudal1}
\end{equation}
The BRST invariance is immediate, just as the $\delta $ invariance since 
\begin{equation}
\delta S_{GF+FP}=s\int d^{4}xc^{A}\partial _{\mu }\mathcal{A}^{A\mu }=\frac{%
s^{2}}{2}\int d^{4}x\mathcal{A}^{2}=0\;.  \label{dudal2}
\end{equation}
It is easy checked that also $\overline{s}$ and $\overline{\delta }$ leave
the action (\ref{dudal1}) invariant. Hence, the NO algebra is a global
symmetry of Yang-Mills theories in the Landau gauge, a fact already longer
known \cite{no,ni}.

\subsection{Curci-Ferrari Gauge}

Next, we discuss Yang-Mills theories in a class of generalized covariant
non-linear gauges proposed in \cite{dj}. The action is given by 
\begin{eqnarray}
S &=&S_{YM}+S_{GF+FP}  \nonumber \\
&=&-\frac{1}{4}\int d^{4}x\mathcal{F}_{\mu \nu }^{A}\mathcal{F}^{A\mu \nu }+s%
\overline{s}\int d^{4}x\left( \frac{1}{2}\mathcal{A}_{\mu }^{A}\mathcal{A}%
^{A\mu }-\frac{\xi }{2}c^{A}\overline{c}^{A}\right)  \nonumber \\
&=&-\frac{1}{4}\int d^{4}x\mathcal{F}_{\mu \nu }^{A}\mathcal{F}^{A\mu \nu
}+\int d^{4}x\left( b^{A}\partial _{\mu }\mathcal{A}^{A\mu }+\frac{\xi }{2}%
b^{A}b^{A}+\overline{c}^{A}\partial ^{\mu }\mathcal{D}_{\mu
}^{AB}c^{B}\right.  \nonumber \\
&-&\left. \frac{\xi }{2}gf^{ABC}b^{A}\overline{c}^{B}c^{C}-\frac{\xi }{8}%
g^{2}f^{ABC}f^{CDE}\overline{c}^{A}\overline{c}^{B}c^{D}c^{E}\right) \;,
\label{cfgf}
\end{eqnarray}
where $\xi $ is the gauge parameter. A gauge fixing as in (\ref{cfgf}) is
sometimes called the Curci-Ferrari (CF) gauge, since its gauge fixing part
resembles the gauge fixing part of the massive, $SU(N)$ gauge model
introduced in \cite{cf}.

In addition to the BRST and anti-BRST symmetries, the action (\ref{cfgf}) is
also invariant under the global $SL(2,R)$ symmetry generated by $\delta $, $%
\overline{\delta }$ \cite{sawa,ds} and $\delta _{FP}$. We conclude that
Yang-Mills theories in the CF gauge have the NO symmetry.

\subsection{Maximal Abelian Gauge}

We decompose the gauge field into its off-diagonal and diagonal parts,
namely 
\begin{equation}
\mathcal{A}_{\mu }=\mathcal{A}_{\mu }^{A}T^{A}=A_{\mu }^{a}T^{a}+A_{\mu
}^{i}T^{\,i},  \label{conn-sun}
\end{equation}
where the index $i$ labels the $N-1$ generators $T^{\,i}$ of the Cartan
subalgebra. The remaining $N(N-1)$ off-diagonal generators $T^{a}$ will be
labelled by the index $a$. Accordingly, the field strength decomposes as 
\begin{equation}
\mathcal{F}_{\mu \nu }=\mathcal{F}_{\mu \nu }^{A}T^{A}=F_{\mu \nu
}^{a}T^{a}+F_{\mu \nu }^{i}T^{\,i}\;,  \label{fsn}
\end{equation}
with the off-diagonal and diagonal parts given respectively by 
\begin{eqnarray}
F_{\mu \nu }^{a} &=&D_{\mu }^{ab}A_{\nu }^{b}-D_{\nu }^{ab}A_{\mu
}^{b}\;+g\,f^{abc}A_{\mu }^{b}A_{\nu }^{c}\;,  \nonumber \\
F_{\mu \nu }^{i} &=&\partial _{\mu }A_{\nu }^{i}-\partial _{\nu }A_{\mu
}^{i}+gf^{abi}A_{\mu }^{a}A_{\nu }^{b}\;\,,  \label{fscompn}
\end{eqnarray}
where the covariant derivative $D_{\mu }^{ab}$ is defined with respect to
the diagonal components $A_{\mu }^{i}$ 
\begin{equation}
D_{\mu }^{ab}\equiv \partial _{\mu }\delta ^{ab}-gf^{abi}A_{\mu
}^{i}\,\,\,\,\,\,.  \label{cdern}
\end{equation}
For the pure Yang-Mills action one obtains 
\begin{equation}
S_{\mathrm{YM}}=-\frac{1}{4}\int d^{4}x\,\left( F_{\mu \nu }^{a}F^{a\mu \nu
}+F_{\mu \nu }^{i}F^{i\mu \nu }\right) \;.  \label{symn}
\end{equation}
The so called MAG gauge condition amounts to fix the value of the covariant
derivative $(D_{\mu }^{ab}A^{b\mu })$ of the off-diagonal components \cite
{th,ks}. However, this condition being nonlinear, a quartic ghost
self-interaction term is required for renormalizability \cite{mp,fz}. The
corresponding gauge fixing term turns out to be \cite{gfk}
\begin{equation}
S_{\mathrm{MAG}}=s\overline{s}\,\int d^{4}x\,\left( \frac{1}{2}A_{\mu
}^{a}A^{a\mu }-\frac{\xi }{2}c^{a}\overline{c}^{a}\right) \;,  \label{smn}
\end{equation}
where $s$ denotes the nilpotent BRST\ operator 
\begin{eqnarray}
&&sA_{\mu }^{a}=-\left( D_{\mu }^{ab}c^{b}+gf^{\,abc}A_{\mu
}^{b}c^{c}+gf^{\,abi}A_{\mu }^{b}c^{i}\right) \,,\hspace{0.3cm}sA_{\mu
}^{i}=-\left( \partial _{\mu }c^{i}+gf\,^{iab}A_{\mu }^{a}c^{b}\right) \;, 
\nonumber \\
&&sc^{a}=gf\,^{abi}c^{b}c^{i}+\frac{g}{2}f\,^{abc}c^{b}c^{c}\,,\hspace{3.1cm}%
sc^{i}=\frac{g}{2}\,f\,^{iab}c^{a}c^{b},  \nonumber \\
&&s\overline{c}^{a}=b^{a}\,,\hspace{6.5cm}s\overline{c}^{i}=b^{i}\;, 
\nonumber \\
&&sb^{a}=0\,,\hspace{6.65cm}sb^{i}=0\;,  \label{BRSTn}
\end{eqnarray}
and $\overline{s}$ the nilpotent anti-BRST\ operator, which acts as 
\begin{eqnarray}
&&\overline{s}A_{\mu }^{a}=-\left( D_{\mu }^{ab}\overline{c}%
^{b}+gf^{\,abc}A_{\mu }^{b}\overline{c}^{c}+gf^{\,abi}A_{\mu }^{b}\overline{c%
}^{i}\right) \,,\hspace{0.3cm}\overline{s}A_{\mu }^{i}=-\left( \partial
_{\mu }\overline{c}^{i}+gf\,^{iab}A_{\mu }^{a}\overline{c}^{b}\right) \;, 
\nonumber \\
&&\overline{s}\overline{c}^{a}=gf\,^{abi}\overline{c}^{b}\overline{c}^{i}+%
\frac{g}{2}f\,^{abc}\overline{c}^{b}\overline{c}^{c}\,,\hspace{3.1cm}%
\overline{s}\overline{c}^{i}=\frac{g}{2}\,f\,^{iab}\overline{c}^{a}\overline{%
c}^{b},  \nonumber \\
&&\overline{s}c^{i}=-b^{i}+gf\,^{ibc}c^{b}\overline{c}^{c}\,,\hspace{4.3cm}%
\overline{s}b^{i}=-gf\,^{ibc}b^{b}\overline{c}^{c}\,,  \nonumber \\
&&\overline{s}c^{a}=-b^{a}+gf\,^{abc}c^{b}\overline{c}^{c}+gf\,^{abi}c^{b}%
\overline{c}^{i}+gf\,^{abi}\overline{c}^{b}c^{i}\,,  \nonumber \\
&&\overline{s}b^{a}=-gf\,^{abc}b^{b}\overline{c}^{c}-gf\,^{abi}b^{b}%
\overline{c}^{i}+gf\,^{abi}\overline{c}^{b}b^{i}\;.\,  \label{aBRSTn}
\end{eqnarray}
Here $c^{a}$ and $c^{i}$ are the off-diagonal and the diagonal components of
the Faddeev-Popov ghost field, $\overline{c}^{a}$ and $\overline{c}^{i}$ the
off-diagonal and the diagonal anti-ghost fields and $b^{a}$ and $b^{i}$ are
the off-diagonal and diagonal Lagrange multipliers. These transformation are
nothing else than the projection on diagonal and off-diagonal fields of (\ref
{s}) and (\ref{as}). Expression $\left( \ref{smn}\right) $ is easily worked
out and yields 
\begin{eqnarray}
S_{\mathrm{MAG}} &=&s\,\int d^{4}x\,\left( \overline{c}^{a}\left( D_{\mu
}^{ab}A^{b\mu }+\frac{\xi }{2}b^{a}\right) -\frac{\xi }{2}gf\,^{abi}%
\overline{c}^{a}\overline{c}^{b}c^{i}-\frac{\xi }{4}gf\,^{abc}c^{a}\overline{%
c}^{b}\overline{c}^{c}\right) \;  \nonumber \\
&=&\int d^{4}x\left( b^{a}\left( D_{\mu }^{ab}A^{b\mu }+\frac{\xi }{2}%
b^{a}\right) +\overline{c}^{a}D_{\mu }^{ab}D^{\mu bc}c^{c}+g\overline{c}%
^{a}f^{abi}\left( D_{\mu }^{bc}A^{c\mu }\right) c^{i}\right.   \nonumber \\
&&\,\,\,\,\,\,\,\,\,\,\,\,\,\,\,\,\,\,+g\overline{c}^{a}D_{\mu }^{ab}\left(
f^{bcd}A^{c\mu }c^{d}\right) -g^{2}f^{abi}f^{cdi}\overline{c}^{a}c^{d}A_{\mu
}^{b}A^{c\mu }-\xi gf^{abi}b^{a}\overline{c}^{b}c^{i}\;  \nonumber \\
&&\,\,\,\,\,\,\,\,\,\,\,\,\,\,\,\,\,-\frac{\xi }{2}gf^{abc}b^{a}\overline{c}%
^{b}c^{c}-\frac{\xi }{4}g^{2}f^{abi}f^{cdi}\overline{c}^{a}\overline{c}%
^{b}c^{c}c^{d}-\frac{\xi }{4}g^{2}f^{abc}f^{adi}\overline{c}^{b}\overline{c}%
^{c}c^{d}c^{i}  \nonumber \\
&&\left. \,\,\,\,\,\,\,\,\,\,\,\,\,\,-\frac{\xi }{8}g^{2}f^{abc}f^{ade}%
\overline{c}^{b}\overline{c}^{c}c^{d}c^{e}\right) \;.  \label{smn2}
\end{eqnarray}
The MAG condition allows for a residual local $U(1)^{N-1}$ invariance with
respect to the diagonal subgroup, which has to be fixed by means of a
suitable further gauge condition on the diagonal components $A_{\mu }^{i}$.
We shall choose a diagonal gauge fixing term which is BRST\ and anti-BRST\
invariant. The diagonal gauge fixing is then given by 
\begin{eqnarray}
S_{\mathrm{diag}} &=&s\overline{s}\,\int d^{4}x\,\;\left( \frac{1}{2}A_{\mu
}^{i}A^{i\mu }\right)   \nonumber \\
&=&s\;\int d^{4}x\,\;\left( \overline{c}^{i}\partial _{\mu }A^{i\mu
}-gf^{iab}A_{\mu }^{i}A^{a\mu }\overline{c}^{b}\right)   \nonumber \\
&=&\int d^{4}x\left( b^{i}\partial _{\mu }A^{i\mu }+\overline{c}^{i}\partial
^{2}c^{i}+gf^{iab}A_{\mu }^{a}(\partial _{\mu }c^{i}\overline{c}%
^{b}-\partial _{\mu }\overline{c}^{i}c^{b})\right.   \nonumber \\
&&\,\,\,\,\,+g^{2}f^{iab}f^{icd}\overline{c}^{a}c^{d}A_{\mu }^{b}A^{c\mu
}-gf^{iab}A_{\mu }^{i}A^{a\mu }(b^{b}-gf^{ibc}\overline{c}^{c}c^{i}) 
\nonumber \\
&&\,\left. \,\,\,+gf^{iab}A^{i\mu }(D_{\mu }^{ac}c^{c})\overline{c}%
^{b}+g^{2}f^{abi}f^{acd}A_{\mu }^{i}A^{c\mu }c^{d}\overline{c}^{b}\right) .
\label{dgf}
\end{eqnarray}
In addition to the BRST\ and the anti-BRST\ symmetry, the gauge-fixed action 
$\left( S_{\mathrm{YM}}+S_{\mathrm{MAG}}+S_{\mathrm{diag}}\right) $ is
invariant under a global $SL(2,R)\,$symmetry, which is generated by the
operators $\delta $, $\overline{\delta }$ and the ghost number operator $%
\delta _{\mathrm{FP}}$. For the $\delta $ transformations we have 
\begin{eqnarray}
&&\delta \overline{c}^{a}=c^{a}\,,\hspace{0.3cm}\delta \overline{c}%
^{i}=c^{i}\,,  \nonumber \\
&&\delta b^{a}=gf^{abi}c^{b}c^{i}+\frac{g}{2}f^{abc}c^{b}c^{c}\,,  \nonumber
\\
&&\delta b^{i}=\frac{g}{2}f^{iab}c^{a}c^{b}\,,  \nonumber \\
&&\delta A_{\mu }^{a}=\delta A_{\mu }^{i}=\delta c^{a}=\delta c^{i}=0\;.
\label{deltan}
\end{eqnarray}
The operator $\overline{\delta }$ acts as 
\begin{eqnarray}
&&\overline{\delta }c^{a}=\overline{c}^{a}\,,\hspace{0.3cm}\overline{\delta }%
c^{i}=\overline{c}^{i}\,,  \nonumber \\
&&\overline{\delta }b^{a}=gf^{abi}\overline{c}^{b}\overline{c}^{i}+\frac{g}{2%
}f^{abc}\overline{c}^{b}\overline{c}^{c}\,,  \nonumber \\
&&\overline{\delta }b^{i}=\frac{g}{2}f^{iab}\overline{c}^{a}\overline{c}%
^{b}\,,  \nonumber \\
&&\overline{\delta }A_{\mu }^{a}=\overline{\delta }A_{\mu }^{i}=\overline{%
\delta }\overline{c}^{a}=\overline{\delta }\overline{c}^{i}=0\;.
\label{deltab}
\end{eqnarray}
The existence of the $SL(2,R)\,$symmetry has been pointed out in \cite{ms}
in the maximal Abelian gauge for the gauge group $SU(2)$. A generalization
of it can be found in \cite{sawa}. There are, however, important differences
between \cite{ms,sawa} and the present analysis.

The first point relies on the choice of the diagonal part of the gauge
fixing $S_{\mathrm{diag}}$, a necessary step towards a complete quantization
of the model. We remark that with our choice of $S_{\mathrm{diag}}$ in eq.$%
\left( \ref{dgf}\right) $, the whole NO algebra becomes an exact symmetry of
the gauge fixed action $\left( S_{\mathrm{YM}}+S_{\mathrm{MAG}}+S_{\mathrm{%
diag}}\right) $ with gauge group $SU(N)$, for any value of $N$. In
particular, as one can see from expression $\left( \ref{dgf}\right) $, $S_{%
\mathrm{diag}}$ contains the interaction term $g^{2}f^{iab}f^{icd}\overline{c%
}^{a}c^{d}A_{\mu }^{b}A^{c\mu }$, which precisely cancels the corresponding
term appearing in eq.$\left( \mathrm{{\ref{smn2}}}\right) $ for $S_{\mathrm{%
MAG}}$. This is a welcome feature, implying that no tachyons are generated
if the $SL(2,R)$, and thus the NO algebra, is required as an exact
invariance for the starting gauge-fixed action. We remark that a similar
compensation holds also for the interaction terms of $\left( \ref{dgf}%
\right) $ and $\left( \mathrm{{\ref{smn2}}}\right) $ containing two diagonal
gluons and a pair of off-diagonal ghost-antighost, implying that the
diagonal gauge fields remain massless.

A second difference concerns the way the fields are transformed. We observe
that in the present case, the field transformations $\left( \ref{BRSTn}%
\right) -\left( \ref{aBRSTn}\right) $ and $\left( \ref{deltan}\right)
-\left( \ref{deltab}\right) $ are obtained from $\left( \ref{s}\right)
-\left( \ref{as}\right) $ and $\left( \ref{df}\right) -\left( \ref{adf}%
\right) $ upon projection of the group index $A=1,...,\left( N^{2}-1\right) $
over the Cartan subgroup of $SU(N)$ and over the off-diagonal generators,
thus preserving the whole NO\ structure. As it is apparent from eqs.$\left( 
\ref{deltan}\right) $, $\left( \ref{deltab}\right) $, the diagonal fields $%
c^{i}$, $\overline{c}^{i}$, $b^{i}$ transform nontrivially, a necessary
feature for the NO algebra. These transformations were not taken into
account in the original work $\,$\cite{ms}. Also, in \cite{sawa}, the NO
structure is analysed only on the off-diagonal fields, the diagonal
components $\overline{c}^{i}$, $b^{i}$ being set to zero.

In summary, it is possible to choose the diagonal part $S_{\mathrm{diag}}$
of the gauge fixing so that the whole NO\ structure is preserved.
Remarkably, the requirement that the NO algebra is an exact symmetry of the
starting action ensures that no tachyons show up. It remains now to prove
that the choice of the diagonal gauge fixing $\left( \ref{dgf}\right) $ will
lead to a renormalizable model. This will be the task of the next section.

\section{Stability of the MAG under radiative corrections}

In order to discuss the renormalizability of the action $\left( S_{\mathrm{YM%
}}+S_{\mathrm{MAG}}+S_{\mathrm{diag}}\right) $ within the BRST\ framework,
we have to write down the Ward identities corresponding to the symmetries of
the classical action. The expression of the BRST\ invariance as a functional
identity requires the introduction of invariant external sources 
\begin{equation}
S_{\mathrm{ext}}=\int d^{4}x\left( \Omega ^{a\mu }sA_{\mu }^{a}+\Omega
^{i\mu }sA_{\mu }^{i}+L^{a}sc^{a}+L^{i}sc^{i}\right) \;,  \label{sext}
\end{equation}
with $s\Omega _{\mu }^{a}=s\Omega _{\mu }^{i}=sL^{a}=sL^{i}=0.$

The $\delta \,$transformations of the external fields can be fixed by
imposing $\delta S_{\mathrm{ext}}=0,$ which yields $\delta \Omega _{\mu
}^{a}=\delta \Omega _{\mu }^{i}=\delta L^{a}=\delta L^{i}=0.$ Therefore the
classical action 
\begin{equation}
\Sigma =S_{\mathrm{YM}}+S_{\mathrm{MAG}}+S_{\mathrm{diag}}+S_{\mathrm{ext}%
}\;,  \label{ta}
\end{equation}
is invariant under BRST\ and $\delta \,$transformations, obeying the
following identities

\begin{itemize}
\item  Slavnov-Taylor identity: 
\begin{eqnarray}
&&\mathcal{S}(\Sigma )=\int d^{4}x\left( \frac{\delta \Sigma }{\delta A_{\mu
}^{a}}\frac{\delta \Sigma }{\delta \Omega ^{a\mu }}+\frac{\delta \Sigma }{%
\delta A_{\mu }^{i}}\frac{\delta \Sigma }{\delta \Omega ^{i\mu }}+\frac{%
\delta \Sigma }{\delta c^{a}}\frac{\delta \Sigma }{\delta L^{a}}\right. 
\nonumber \\
&&\left.
\,\,\,\,\,\,\,\,\,\,\,\,\,\,\,\,\,\,\,\,\,\,\,\,\,\,\,\,\,\,\,\,\,\,\,\,+%
\frac{\delta \Sigma }{\delta c^{i}}\frac{\delta \Sigma }{\delta L^{i}}+b^{a}%
\frac{\delta \Sigma }{\delta \overline{c}^{a}}+b^{i}\frac{\delta \Sigma }{%
\delta \overline{c}^{i}}\right) =0\;  \label{sti}
\end{eqnarray}

\item  $\delta $ symmetry Ward identity: 
\begin{equation}
\mathcal{D}(\Sigma )=\int d^{4}x\left( c^{a}\frac{\delta \Sigma }{\delta 
\overline{c}^{a}}+c^{i}\frac{\delta \Sigma }{\delta \overline{c}^{i}}+\frac{%
\delta \Sigma }{\delta b^{a}}\frac{\delta \Sigma }{\delta L^{a}}+\frac{%
\delta \Sigma }{\delta b^{i}}\frac{\delta \Sigma }{\delta L^{i}}\right) =0
\label{dwi}
\end{equation}

\item  Integrated diagonal ghost equation: 
\begin{equation}
\mathcal{G}^{i}\Sigma =\Delta _{\mathrm{cl}}^{i}\;,  \label{gen}
\end{equation}
where 
\begin{equation}
\mathcal{G}^{i}\mathcal{=}\int d^{4}x\left( \frac{\delta }{\delta c^{i}}%
+gf\,^{abi}\overline{c}^{a}\frac{\delta }{\delta b^{b}}\right)  \label{ge}
\end{equation}
and the classical breaking $\Delta _{\mathrm{cl}}^{i}$ is given by 
\begin{equation}
\Delta _{\mathrm{cl}}^{i}=\int d^{4}x\left( gf\,^{abi}\Omega _{\mu
}^{a}A^{b\mu }-gf\,^{abi}L^{a}c^{b}\right) \;.  \label{cbn}
\end{equation}
\end{itemize}

\begin{itemize}
\item  Integrated diagonal antighost equation:
\end{itemize}

\begin{equation}
\int d^{4}x\frac{\delta \Sigma }{\delta \overline{c}^{i}}=0\;.  \label{age}
\end{equation}
Similarly, we could also impose the anti-BRST\ and $\overline{\delta }$
symmetries in a functional way by introducing an additional set of external
sources. However, the Ward identities $\left( \ref{sti}\right) -\left( \ref
{age}\right) $ are sufficient to ensure the stability of the classical
action under quantum corrections. It is not difficult indeed, by using the
algebraic renormalization procedure \cite{book,bbh}, to prove that the model
is renormalizable.

\section{Ghost condensation and the breakdown of $SL(2,R)$ and NO symmetry}

In this section, we give a brief discussion of the existence of ghost
condensates and of their relationship with $SL(2,R)$ and hence with NO
symmetry. \newline
\newline
The condensation of ghosts came to attention originally in the works of \cite
{ms,k} in the context of $SU(2)$ MAG. The decomposition of the 4-ghost
interaction allowed to construct an effective potential with a nontrivial
minimum for the off-diagonal condensate $\left\langle \varepsilon ^{3ab}c^{a}%
\overline{c}^{b}\right\rangle $. This condensate implies a dynamical
breaking of $SL(2,R)$. Order parameters for this breaking are given by

\begin{equation}
\left\langle \varepsilon ^{3ab}c^{a}\overline{c}^{b}\right\rangle =\frac{1}{2%
}\left\langle \delta \left( \varepsilon ^{3ab}\overline{c}^{a}\overline{c}%
^{b}\right) \right\rangle =\frac{1}{2}\left\langle \overline{\delta }\left(
\varepsilon ^{3ab}c^{a}c^{b}\right) \right\rangle \;.  \label{op}
\end{equation}
As a consequence the generators $\delta $ and $\overline{\delta }$ of $%
SL(2,R)$ are broken \cite{ms}, while the Faddeev-Popov ghost number
generator $\delta _{FP}=\left[ \delta ,\overline{\delta }\right] $ remains
unbroken. The ghost condensate resulted in a mass for the off-diagonal
gluons, whose origin can be traced back to the presence of the off-diagonal
interaction term $\overline{c}cA_{\mu }A^{\mu }$ term in expression (\ref
{smn2}). However, as was shown in \cite{dd}, this mass is a tachyonic one.
Furthermore, the requirement of invariance of $S_{diag}$ under anti-BRST
and, as a consequence, under $SL(2,R)$ transformations, also gives rise to
quartic terms of the kind $\overline{c}cA_{\mu }A^{\mu }$ (see (\ref{dgf})),
which precisely cancel those of $S_{MAG}$ in (\ref{smn2}). Thus, if the $%
SL(2,R)$ symmetry (NO symmetry (\ref{ojima})) is required, the ghost
condensates do not induce any unphysical mass for the off-diagonal gluons.%
\newline
\newline
The 4-ghost interaction can be decomposed in a different way, so that the
ghost condensation takes places in the Faddeev-Popov charged channels $%
\varepsilon ^{3ab}c^{a}c^{b}$ and $\varepsilon ^{3ab}\overline{c}^{a}%
\overline{c}^{b}$ instead of $\varepsilon ^{3ab}c^{a}\overline{c}^{b}$. In
this case, the ghost number symmetry is broken \cite{sp}. Consequently, the
NO algebra is broken again. It is interesting to see that the existence of
different ghost channels in which the ghost condensation can take place has
an analogy in ordinary superconductivity, known as BCS (particle-particle
and hole-hole pairing) versus Overhauser (particle-hole paring) \cite{bo,bo1}%
. In the present case the BCS\ channel corresponds to the Faddeev-Popov
charged condensates $\left\langle \varepsilon ^{3ab}c^{a}c^{b}\right\rangle $
and $\left\langle \varepsilon ^{3ab}\overline{c}^{a}\overline{c}%
^{b}\right\rangle $, while the Overhauser channel to $\left\langle
\varepsilon ^{3ab}c^{a}\overline{c}^{b}\right\rangle $. \newline
\newline
The CF gauge (\ref{cfgf}) contains a 4-ghost interaction too, so it is
expected that the ghost condensation can take place also in the CF gauge.
This was confirmed in \cite{sp2}. Notice that the CF and MAG gauges look
very similar, and since the CF gauge does not contain terms like $\overline{c%
}c\mathcal{A}_{\mu }\mathcal{A}^{\mu }$, no (tachyonic) mass is induced for
the gluons. This can be seen as some extra evidence why also in the MAG no
tachyon mass terms should be generated.\newline
\newline
More surprising is the fact that also in the Landau gauge, the ghost
condensation occurs \cite{sp3}. Since there is no 4-ghost interaction to be
decomposed, another technique was used to discuss this gauge. A combination
of the algebraic renormalization technique \cite{book,bbh} and the local
composite operator technique \cite{h1} allowed a clean treatment, with the
result that also in case of the Landau gauge the NO symmetry is broken. 
\newline
\newline
Since the aforementioned ghost condensates are not giving masses for the
gluons, one could wonder what the mechanism behind the dynamical generation
of gluon masses could be. It was proposed in \cite{ope} that the generation
of a real mass for the gluons in case of the CF gauge is related to a
non-vanishing vacuum expectation value for the two-dimensional local,
composite operator $\left( \frac{1}{2}\mathcal{A}_{\mu }^{A}\mathcal{A}%
^{A\mu }-\xi c^{A}\overline{c}^{A}\right) $. It is interesting to notice
that this is exactly the kind of mass term that is present in the massive
Lagrangian of Curci and Ferrari \cite{cf}. In the case of MAG, the relevant
operator is believed to be $\left( \frac{1}{2}A_{\mu }^{a}A^{a\mu }-\xi c^{a}%
\overline{c}^{a}\right) $, and is expected to provide an effective mass for
both off-diagonal gauge and off-diagonal ghost fields \cite{ope,dd}.

\section{Conclusion}

\begin{itemize}
\item  In this paper the presence of the $SL(2,R)$ symmetry has been
analysed in the Landau, Curci-Ferrari and maximal Abelian gauge for $SU(N)$
Yang-Mills. In all these gauges $SL(2,R)$ can be established as an exact
invariance of the complete gauge fixed action. Together with the BRST\ and
anti-BRST, the generators of $SL(2,R)$ are part of a larger algebra, known
as the Nakanishi-Ojima algebra \cite{no}.

\item  In particular, we have been able to show that in the case of the
maximal Abelian gauge, the requirement of $SL(2,R)$ for the complete action,
including the diagonal gauge fixing term, ensures that no tachyons will be
generated.

\item  In all these gauges, the $SL(2,R)$ symmetry turns out to be
dynamically broken by the existence of off-diagonal ghost condensates $%
\left\langle cc\right\rangle $, $\left\langle \overline{c}\overline{c}%
\right\rangle $ and $\left\langle c\overline{c}\right\rangle $. As a
consequence, the NO algebra is also broken. These condensates deeply modify
the infrared behavior of the off-diagonal ghost propagator, while
contributing to the vacuum energy density and hence to the trace anomaly 
\cite{ms,k,sp}. \newline

\item  Finally, let us spend a few words on future research. As already
remarked, the ghost condensation can be observed in different channels,
providing a close analogy with the BCS\ versus Overhauser effect of
superconductivity. We have also pointed out that the existence of the
condensate $\left( \frac{1}{2}\left\langle A_{\mu }^{a}A^{a\mu
}\right\rangle -\xi \left\langle c^{a}\overline{c}^{a}\right\rangle \right) $
can be at the origin of the dynamical mass generation in the MAG for all
off-diagonal gluons and ghosts \cite{ope,dd}, a feature of great relevance
for the Abelian dominance. Both aspects will be analysed by combining the
algebraic renormalization \cite{book,bbh} with the local composite operator
technique \cite{h1}, as done in the case of the ghost condensation in the
Landau gauge \cite{sp3}. The combination of these two procedures results in
a very powerful framework for discussing the ghost condensation in the
various channels as well as for studying the condensate $\left( \frac{1}{2}%
\left\langle A_{\mu }^{a}A^{a\mu }\right\rangle -\xi \left\langle c^{a}%
\overline{c}^{a}\right\rangle \right) $ and its relationship with the
dynamical mass generation. Also, the detailed analysis of the decoupling at
low energies of the diagonal ghosts and of the validity of the local $%
U(1)^{N-1}$ Ward identity in the MAG deserves careful attention.
\end{itemize}

\section*{Acknowledgments}

The Conselho Nacional de Desenvolvimento Cient\'{i}fico e Tecnol\'{o}gico
CNPq-Brazil, the Funda{\c{c}}{\~{a}}o de Amparo a Pesquisa do Estado do Rio
de Janeiro (Faperj), the SR2-UERJ and the MIUR-Italy are acknowledged for
the financial support.


\begin{thebibliography}{99}
\bibitem{scon}  Y. Nambu, \emph{Phys. Rev. }\textbf{D10} (1974) 4262;\newline
G. 't Hooft, \emph{High Energy Physics EPS Int. Conference, }Palermo 1975,
ed. A. Zichichi;\newline
S. Mandelstam, \emph{Phys. Rept. }\textbf{23} (1976) 245.

\bibitem{th}  G. 't Hooft, \emph{Nucl. Phys. }\textbf{B190} [FS3] (1981) 455.

\bibitem{ks}  A. Kronfeld, G. Schierholz and U.-J. Wiese, \emph{Nucl. Phys. }%
\textbf{B293} (1987) 461;\newline
A. Kronfeld, M. Laursen, G. Schierholz and U.-J. Wiese, \emph{Phys. Lett. }%
\textbf{B198} (1987) 516.

\bibitem{mp}  H. Min, T. Lee and P.Y. Pac, \emph{Phys. Rev. }\textbf{D32}
(1985) 440.

\bibitem{fz}  A.R. Fazio, V.E.R. Lemes, M.S.\ Sarandy and S.P. Sorella, 
\emph{Phys. Rev. }\textbf{D64} (2001) 085003.

\bibitem{ms}  M. Schaden, \emph{Mass Generation in Continuum SU(2) Gauge
Theory in Covariant Abelian Gauges}, hep-th/9909011; \emph{Mass Generation,
Ghost Condensation and Broken Symmetry: SU(2) in Covariant Abelian Gauges},
talk given at Confinement IV, Vienna, 2000, hep-th/0108034; \emph{SU(2)
Gauge Theory in Covariant (Maximal) Abelian Gauges}, talk presented at Vth
Workshop on QCD, Villefranche, 2000, hep-th/0003030.

\bibitem{k}  K.-I. Kondo and T. Shinohara, \emph{Phys. Lett. }\textbf{B491}
(2000) 263.

\bibitem{sp}  V.E.R. Lemes, M.S.\ Sarandy and S.P. Sorella, \emph{Ghost
Number Dynamical Symmetry Breaking in Yang-Mills Theories in the Maximal
Abelian Gauge}, hep-th/0206251.

\bibitem{dd}  D. Dudal and H. Verschelde, \emph{On ghost condensation, mass
generation and Abelian dominance in the Maximal Abelian Gauge, }%
hep-th/0209025.

\bibitem{sp2}  A.R. Fazio, V.E.R. Lemes, M. Picariello, M.S.\ Sarandy and
S.P. Sorella, \emph{Ghost condensates in Yang-Mills theories in nonlinear
gauges}, hep-th/0210036.

\bibitem{sp3}  V.E.R. Lemes, M.S.\ Sarandy and S.P. Sorella, \emph{Ghost
condensates in Yang-Mills theories in the Landau gauge}, hep-th/0210077.

\bibitem{no}  K. Nakanishi and I. Ojima, \emph{Z. Phys. }\textbf{C6}(1980)
155

\bibitem{ni}  K. Nishijima, \emph{Prog. Theor. Phys. }\textbf{72 }(1984)
1214; \textbf{73} (1985) 536.

\bibitem{dj}  R. Delbourgo and P.D. Jarvis, \emph{J. Phys. }\textbf{A15}
(1982) 611;

L. Baulieu and J. Thierry-Mieg, \emph{Nucl. Phys. }\textbf{B197 }(1982) 477.

\bibitem{oji}  I. Ojima, \emph{Z.Phys. }\textbf{\ C13} (1982) 173.

\bibitem{cf}  G. Curci and R. Ferrari, \emph{Nuovo Cim. }\textbf{A32 }(1976)
151; \emph{Phys. Lett. }\textbf{B63} (1976) 91.

\bibitem{ds}  F. Delduc and S.P.\ Sorella, \emph{Phys. Lett. }\textbf{B231}
(1989) 408.

\bibitem{ope}  K.-I. Kondo, \emph{Phys. Lett.} \textbf{B514} (2001) 335;
K.-I.Kondo, T.Murakami, T.Shinohara, T.Imai, \emph{Phys. Rev.} \textbf{D65}
(2002) 085034.

\bibitem{h1}  H. Verschelde, K. Knecht, K. Van Acoleyen and M. Vanderkelen, 
\emph{Phys. Lett.} \textbf{B516 }(2001) 307.

\bibitem{pb}  Ph. Boucaud, J.P. Leroy, A. Le Yaouanc, J. Micheli, O.
P\`{e}ne, F. De Soto, A. Donini, H. Moutarde, J. Rodr\'{i}guez-Quintero, 
\emph{Phys. Rev.}\textbf{\ D66 (}2002) 034504.

\bibitem{sawa}  H. Sawayanagi, \emph{Prog. Theor. Phys. }\textbf{106 }(2001)
971.

\bibitem{gfk}  K.-I. Kondo, \emph{Phys. Rev.} \textbf{D58} (1998) 105019.

\bibitem{book}  O. Piguet and S.P. Sorella, \emph{Algebraic Renormalization}%
, Monograph series \textbf{m28}, Springer Verlag, 1995.\emph{\ }

\bibitem{bbh}  G. Barnich, F. Brandt and M. Henneaux, \emph{Phys. Rept.} 
\textbf{338 }(2000) 439.

\bibitem{bo}  A. W. Overhauser, \emph{Advances in Physics }\textbf{27 }%
(1978) 343.

\bibitem{bo1}  B.-Y. Park, M. Rho, A. Wirzba, I. Zahed, \emph{Phys. Rev. }%
\textbf{D62} (2000) 034015.
\end{thebibliography}
\end{document}